\documentclass[aps,preprint]{revtex4}
\usepackage{epsfig}
\usepackage{amsmath}
\usepackage{epsfig}

\begin{document}
\title
{\bf Revisiting double Dirac delta potential}    
\author{Zafar Ahmed$^1$, Sachin Kumar$^2$, Mayank Sharma$^3$, Vibhu Sharma$^{3,4}$}
\affiliation{$~^1$Nuclear Physics Division, Bhabha Atomic Research Centre, Mumbai 400085, India\\ 
$~^2$Theoretical Physics Division, Bhabha Atomic Research Centre, Mumbai 400085, India\\
$^{3,4}$Amity Institute of Applied Sciences, Amity University, Noida, UP, 201313, India}
\email{1:zahmed@barc.gov.in, 2: Sachinv@barc.gov.in, 3: mayank.edu002@gmail.com, 4:svibhu876@gmail.com}
\date{\today}
\begin{abstract}
\noindent
We study a general double Dirac delta  potential to show that this is the simplest yet versatile solvable potential to introduce double wells, avoided crossings,  resonances  and perfect transmission ($T=1$). Perfect transmission energies turn out to be the critical property of symmetric and anti-symmetric cases wherein these discrete energies are found to correspond to the eigenvalues of Dirac delta potential placed symmetrically between two rigid walls. For  well(s) or barrier(s), perfect transmission [or zero reflectivity, $R(E)$] at energy $E=0$ is non-intuitive. However, earlier this has been found and called ``threshold anomaly". Here we show that it is  a critical phenomenon and we can have $ 0 \le R(0)<1$ when the parameters of the double delta potential satisfy an interesting condition. We also invoke zero-energy and zero curvature eigenstate ($\psi(x)=Ax+B$) of delta well between two symmetric rigid  walls  for $R(0)=0$.  We  resolve that the resonant energies and the perfect transmission energies are different
and they arise differently. 
\end{abstract}
\maketitle
\section{Introduction}
The general one-dimensional Double Dirac Delta Potential (DDDP) is written as
[see Fig. 1(a,b,c)]
\begin{equation}
V(x)=V_1 \delta(x+b) + V_2 \delta (x-a).
\end{equation} 
When $V_1=V_2=-\alpha$ and $b=a$ it becomes symmetric double delta potential [1-3] which is well known to have at most two discrete eigenvalues; one when $\frac{\hbar^2}{2ma\alpha}>1$ and two when $\frac{\hbar^2}{2ma\alpha}<1$. The symmetric DDDP has also been studied [4] as a scattering potential possessing oscillatory transmission coefficient $T(E)$ as a function of energy.  Using the potential (1), a subtle ``threshold anomaly'' in the scattering from one-dimensional attractive potential wells has been revealed earlier [5]. According to this  the reflection probability becoming anomalous ($R(E=0)=0$) is directly related to the fact whether the potential is at the threshold of possessing a bound state near $E=0$. Here we show a critical nature of this effect. This attractive double delta potential has also been studied for an interesting effect that the Wigner's time-delay [6] at small energies is very large [7] if the potential supports a bound state near $E=0$ or if its strength is just enough to support another bound state. 

Notwithstanding the versatility and simplicity of this potential (1), it has not been  utilized fully in textbooks. Here in this paper, we show that this is the simplest potential to introduce double wells, the rare avoided crossing of two levels in one-dimension, perfect transmission and resonances. Also, in the textbooks, the utility of transmission ($\tau$) and reflection ($\rho$) amplitudes  in extracting bound states, resonances and perfect transmission energies is not often  discussed [2,3,6,9,10]. Here in this paper, we first explain these connections in general  and demonstrate the extraction of three discrete energy spectra of: bound states, resonances and perfect transmission from $R=|\rho|^2$ and $T=|\tau|^2$ for the general double Dirac delta potential given as (1).
\begin{figure}[ht]
\centering
\includegraphics[width=4 cm,height=4 cm]{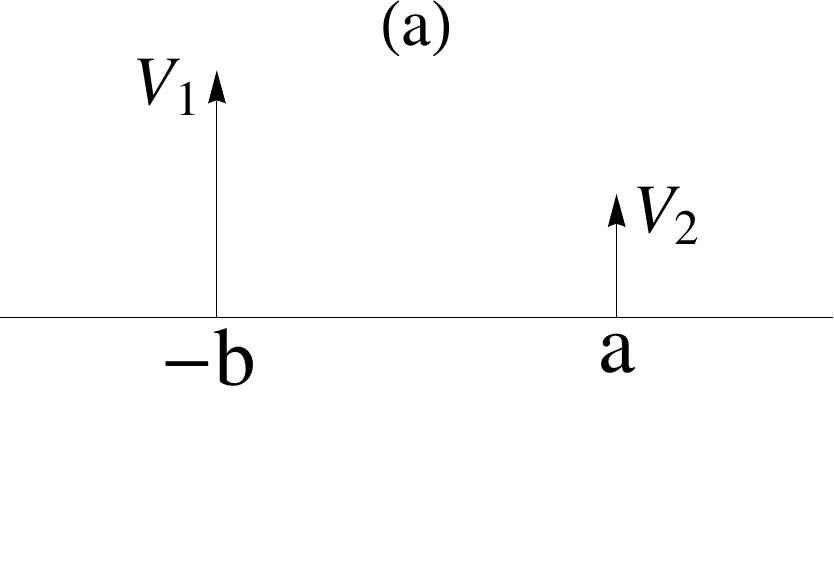}
\hskip .5 cm
\includegraphics[width=4 cm,height=4 cm]{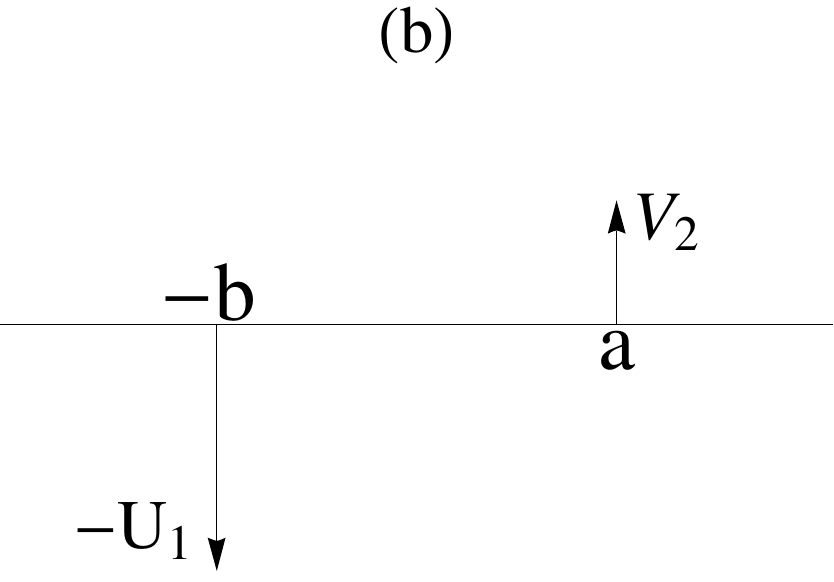}
\hskip .5 cm
\includegraphics[width=4 cm,height=4 cm]{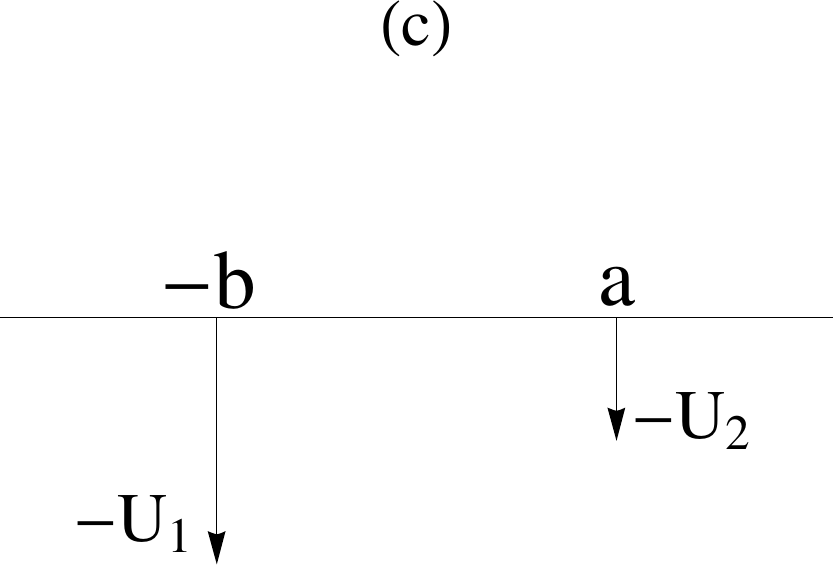}\\ \vskip .5 cm
\includegraphics[width=4 cm,height=4 cm]{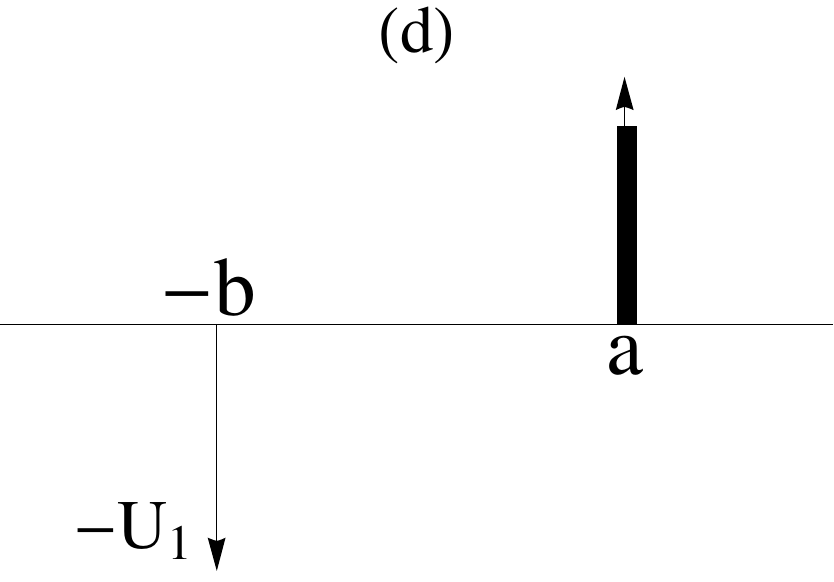}
\hskip .5 cm
\includegraphics[width=4 cm,height=4 cm]{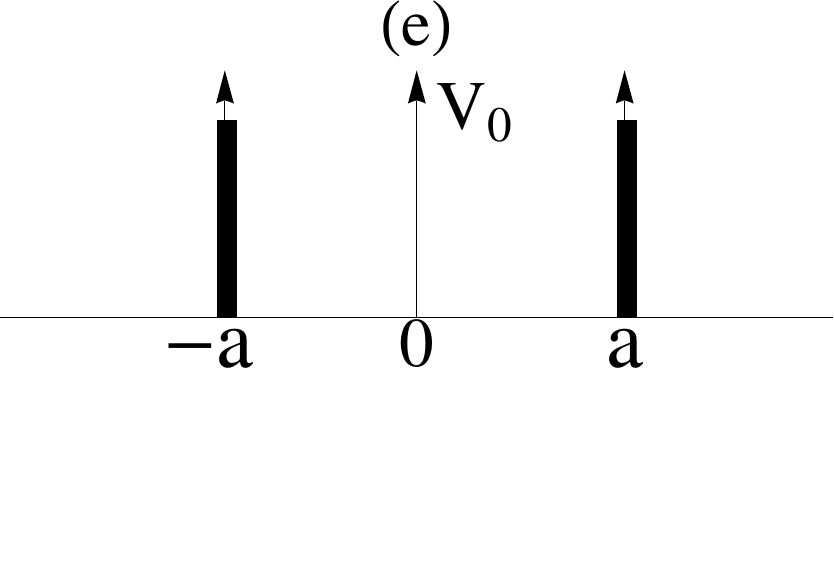}
\hskip .5 cm
\includegraphics[width=4 cm,height=4 cm]{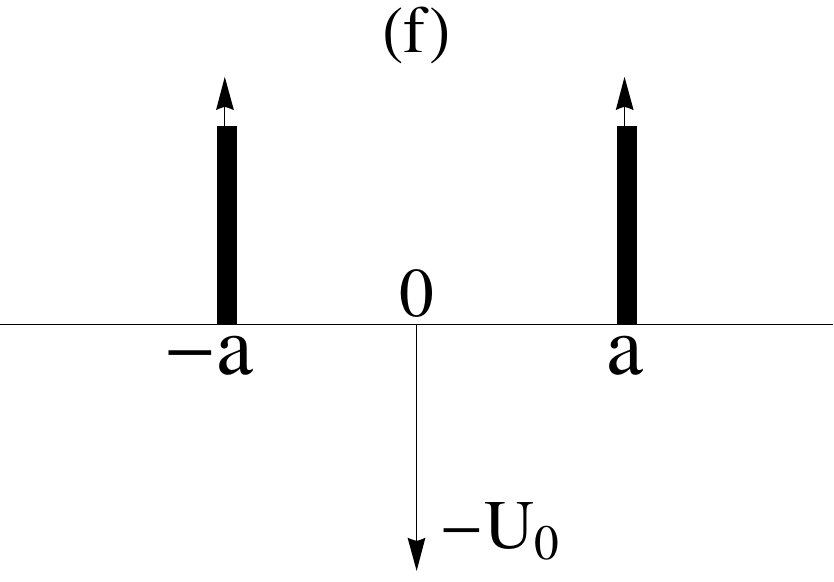}
\caption{(a-d):Depiction of various cases of the double Dirac delta potential (1). In (e) and (f), see the hard-box potentials, where the Dirac delta barrier or  well has been placed symmetrically between two rigid walls, respectively.}
\end{figure}
\section{Scattering coefficients and discrete energies}
The one-dimensional, time-independent Schr{\"o}dinger equation is written as
\begin{equation}
\frac{d^2\psi(x)}{dx^2}+\frac{2m}{\hbar^2}[E-V(x)]\psi(x)=0.
\end{equation}
Let us define 
\begin{equation}
k=\sqrt{\frac{2mE}{\hbar^2}},~ E>0; \quad p=\sqrt{\frac{-2mE}{\hbar^2}}, ~E<0.
\end{equation}
Consider a particle-wave incident on the potential from far left ($x<0$) where $V(x)=0$, it may be reflected back towards left $(x<0)$ and transmitted towards far right  $(x>a)$, where the $V(x)$ is zero again. Every textbook [2,3,6,9-10] writes the general solution of $\psi(x)$ in this case as
\begin{equation}
\psi(x<0)= A e^{ikx} + B e^{-ikx}, \quad \psi(x>a)= F e^{ikx}.
\end{equation}
For the region $0<x<a$, the particular solution of Schr{\"o}dinger equation (a combination of two linearly independent solutions) for the given potential, is juxtaposed between these two solutions.
One then matches the wave function and its derivative at $x=0,a$ to obtain the reflection and transmission amplitudes usually as
\begin{equation}
\rho=\frac{B}{A}, \quad \tau=\frac{F}{A},
\end{equation}
where $A,B,F$ are functions of energy and mass of the particle and the potential parameters. By reversing the signs of the strengths of the potential barrier (e.g., $V_1,V_2$ in (1) we can get $\rho'=B'/A'$ and $\tau'=F'/A'$ for the well. Let us change $k$ to $ip$
in the equation {\it a la} (4). So on the left, we have $\psi(x<-a)= A' e^{-px} + B' e^{px}$ and 
$\psi(x>a)= F' e^{-px}$  on the right. In order to have bound states
we demand  the  wave function  to converge to zero as $x\rightarrow \pm \infty$. This requires $A'=0$ at $k=ip_n, p_n>0$, which in turn are the poles of $\rho'$ and $\tau'$ at negative energies, $E_n=\frac{\hbar^2p_n^2}{2m} <0$. One may also find  the negative energy poles of  $T'(E)$ and $R'(E)$. We would like to caution that in Ref. [6] (on page 109 in Fig. 6.9 ) the negative energy poles  mistakenly appear as spikes of height 1 in the graph of $T(E).$ 

Students could be further instructed to find the poles of  $\rho$ and $\tau$ that amounts to finding the zeros of $A$, this turns the solutions (4) into Gamow's pioneering idea [8,9] of out-going-wave boundary condition at the exit of the potential
\begin{equation}
\psi(x<0)= B e^{-ikx}, \quad \psi(x>a)=F e^{ikx}.
\end{equation}
In case the potential possesses resonances, one gets the poles at $E={\cal E}_n-i\Gamma_n/2$ or $k={\cal K}_n-ik_n'$. The solution $\Psi(x,t)=\psi(x) e^{-iEt/\hbar}$ of the time dependent Schr{\"o}dinger equation for the resonant state can be written as 
\begin{equation}
\Psi(x>a,t)= F ~ e^{i{\cal K}_n x}~ e^{k_n'x} ~ e^{-i{\cal E}_n t/\hbar} ~ e^{-\Gamma_n t/2\hbar},
\end{equation}
where ${\cal K}_n, k'_n >0$, so that $\Gamma_n= = 4{\cal K}_n k'_n >0$.
One can see that the spatial part is an oscillating wave with growing amplitude (spatial catastrophe) for $x\rightarrow \infty$. Similarly one
can get spatial catastrophe on the left side $x \rightarrow -\infty.$ Time-wise,  $\Psi(x,t)$  is  well known as Gamow's decaying state.
These states have explained  then enigmatic phenomenon of alpha-decay from nucleus [8]. Recalling that Hermitian Hamiltonians have real eigenvalues is misplaced here as we are not imposing the condition of bound state  that $\psi(x<0)= e^{px}, \psi(x>a)=e^{-px}$, where $p=\sqrt{-2mE}/\hbar$. Also the spatial catastrophe $e^{k'_n x}$ in (7) will be controlled [9] by the time-wise decaying part $e^{-\Gamma_n t/2\hbar}$. 
\section{The versatile double Dirac delta potential}
In the following we obtain the $\rho$ and $\tau$ from scattering states of the potential (1) ($b=0$ in Fig. 1(a,b,c)), we write the plane wave solution of (2) in different regions as
\begin{equation}
\psi(x<0)= A e^{ikx}+ Be^{-ikx}, ~ \psi(0<x<a) = C e^{ikx} +D e^{-ikx},~ \psi(x>a)= F e^{ikx},\quad  v_j=\frac{2mV_j}{\hbar^2},
\end{equation}
as the potential (1) is zero excepting at two points $x=0,a$ where it is suddenly infinite and hence discontinuous. Normally, the solution of Schr{\"o}dinger equation needs to be both continuous and differentiable. 
However, if a potential has the Dirac delta discontinuity (say) at $x=c$, such that $V(x)=\tilde V(x)+ P \delta(x-c)$, where $\tilde V(x)$ 
is continuous at $x=c$, the integration of (2) from $x=c-\epsilon$ to $x=c+\epsilon$ and then the limit as $\epsilon \rightarrow 0$ yields
\begin{equation}
\left .\frac{d\psi(x)}{dx} \right |_{x<c}-\left .\frac{d\psi(x)}{dx} \right |_{x>c}= \frac{2mP}{\hbar^2} \psi(c).
\end{equation}
This is the well known condition of momentum mismatch
due to Dirac Delta function [2,3,9,10] at
$x=c$. So the condition of continuity and (9) at $x=0,a$ gives the following equations.
\begin{eqnarray}
A+B=C+D, \quad   ik[(C-D)-(A-B)]=v_1 (A+B) \\ \nonumber
C e^{ika} +D e^{-ika} = F e^{ika}, \quad ik [F e^{ika}- C e^{ika}+ D e^{-ika}]=v_2 Fe^{ika}.
\end{eqnarray}
By using these equations we obtain the reflection and transmission 
amplitudes for the potential (1) as
\begin{equation}
\rho=\frac{B}{A}=\frac{2ik(v_1e^{-ika}+v_2e^{ika})+2iv_1v_2 \sin ka}{(2ik-v_1)(2ik-v_2)e^{-ika}-v_1 v_2 e^{ika}} 
\end{equation}
\begin{equation}
\tau=\frac{F}{A}=\frac{-4k^2e^{-ika}}{(2ik-v_1)(2ik-v_2)e^{-ika}-v_1 v_2 e^{ika}}\\
\end{equation}
We now propose to  extract various discrete spectra from Eqs. (11,12).
\subsection{Bound states:}
Let us change in (11,12) $V_1, V_2, k \rightarrow  -U_1, -U_2, ip$ which amounts to changing $v_j$ (8) to $u_j$ (see (13) below). The poles of $\rho$ and $\tau$ are then given by
\begin{equation}
(2p-u_1)(2p-u_2)=u_1u_2 e^{-2pa}, \quad u_j=\frac{2m U_j}{\hbar^2}.
\end{equation}   
For finding the bound state eigenvalues of the double delta wells (Fig. 1(c)) one has to solve this implicit equation numerically. Since this one-dimensional finite potential {\it well} satisfies the condition that $|\int_{-\infty}^{\infty} V(x) dx| <\infty$ (finite) [11], it will have at least one bound state eigenvalue. Next, on the left of the  Eq. (13) we have a quadratic (parabolic) function of $p$ and on the right we have a decreasing exponential, consequently they can cut each other at most at two values of $p$. So there can be at most two discrete eigenvalues in the double delta well (Fig. 1(c)). Let us first recover the well known results in the special cases.

When $a=0$ this potential becomes a single delta well at $x=0$ with the strength as $(U_1+U_2)$ in this case from (13), we get 
\begin{equation}
E=-\frac{m}{\hbar^2} \frac{(U_1+U_2)^2}{2}.
\end{equation} 
Further if $U_2=0$, we get the well known single eigenvalue of the Dirac delta potential as $E=-\frac{mU_1^2}{2\hbar^2}$ [2,3,9,10]. Next when $U_1=U_0, U_2=-U_0$, from (13) we get the eigenvalue equation in this case as
\begin{equation}
(4p^2-u_0^2)= -u_0^2 e^{-2pa} < 0 \Rightarrow E > -\frac{m U_0^2}{2\hbar^2}.
\end{equation}
If we take $u_2 \rightarrow \infty$, the potential (1) becomes a delta well near a rigid wall (Fig. 1(d)) which is a well studied potential [2,3]. For this case let us divide Eq.(13) by $u_2$ and take the limit $u_2 \rightarrow \infty$ we get the eigenvalue equation as
\begin{equation}
e^{-2pa}=1-2p/u_1,
\end{equation} 
the single  eigenvalue will occur only if $u_1a>1$.
When $u_2$ is changed from 0 to $\infty$ the single eigenvalue of (1) (Fig. 1(d))
will vary from $E=-\frac{mV_1^2}{2\hbar^2}$ to the root of the Eq. (16).
So if the delta well is strong enough ($u_1a>1$) even the rigid wall 
perturbation at $x=a$ near the delta well cannot remove the single bound level from the well. 

Again if $U_1=U_0=U_2$ but $a\ne 0$, the whole expression in (13) gets factored into two well known equations [2,3]
\begin{equation}
e^{-pa}=2p/u_0 -1>0, \quad e^{-pa}=1-2p/u_0>0
\end{equation}
The first of these Eqs. (17) always  has a real root for any positive value of $u_0$ confirming an unconditional bound state in the potential. 
The second equation above  will have one real root only when $u_0 a>2$ [2,3], so the first excited state  exists  conditionally. 

Notice that $p=0$ is an unconditional root of (13), so $E=0$ can be mistaken to be an essential bound state of (1) (Fig. 1(c)). Let us investigate Eq. (13)  for $p \approx 0$, ignoring $p^2$ and writing $e^{-pa} \approx 1-pa$, we get 
\begin{equation}
\frac{1}{u_1}+\frac{1}{u_2}=a,
\end{equation}
meaning that when $a>1/u_1+1/u_2$, the first excited state $E_1<0$ would start appearing near $E=0$ in the potential (1) (Fig. 1(c)). When $u_1=u_0=u_2$ this condition becomes  $u_0a>2$ see Fig (2a) for $u_0=11$, $E_1$ starts appearing when $a>2/11$. In Fig. (2b), $u_1=11, u_2=12$, notice that $E_1$ starts appearing when $a>23/132$.
\begin{figure}[ht]
\centering
\includegraphics[width=7 cm,height=7 cm]{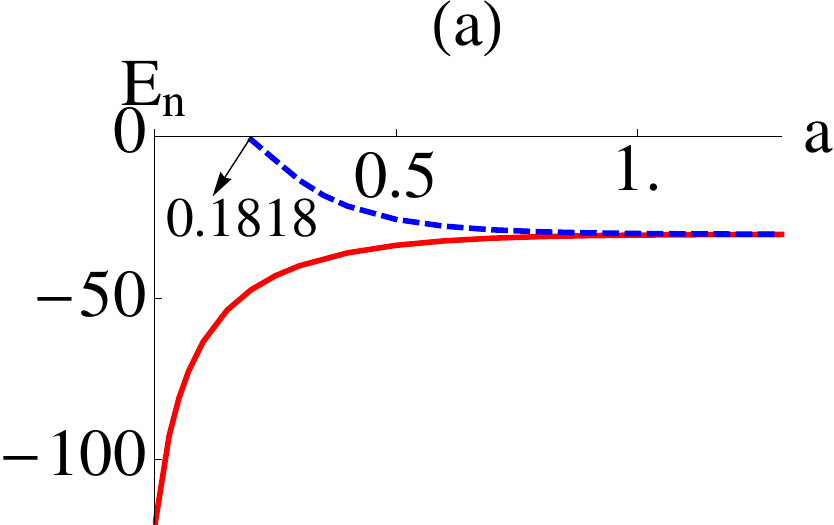}
\hskip .5 cm
\includegraphics[width=7 cm,height=7 cm]{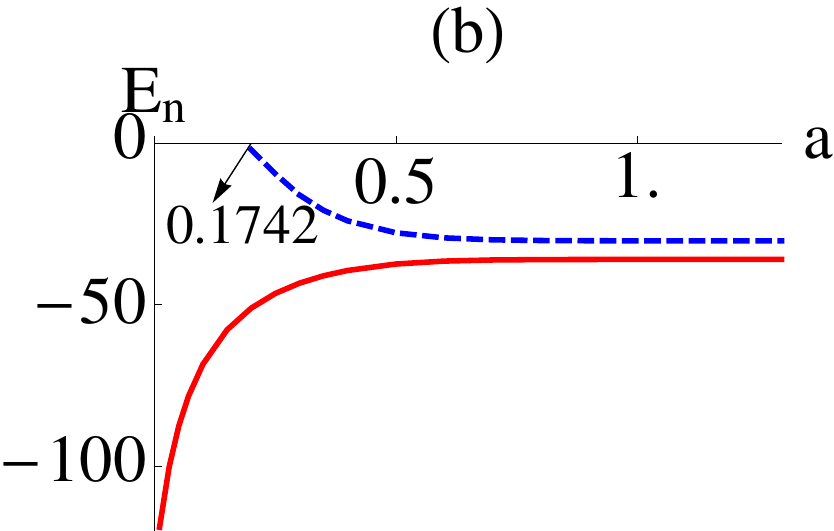}
\caption{ The variation of two levels of double well (Fig.1(c)) potential (1) as the distance between the well, $a$, is varied when (a) $u_1=11=u_2$ (b) $u_1=11, u_2=12$. In (a) the level $E_1$ starts appearing for $a>2/11=0.1818$ and in (b) it appears for $a>23/132=0.1742$ as per Eq. (18). In (a) two levels merge to one level at $E=-30.25$ whereas in (b) two levels saturate to $E=-36$ and $E=-30.25$ (ground state eigenvalues of two independent delta potentials with depths as 12 and 11, respectively. Here and in all the figures below we have taken $2m=1=\hbar^2.$}
\end{figure}
In Fig. 2, we show that characteristic double well behaviour [6] of the potential (1) (Fig. 1(c)), in the symmetric case (a) $u_1=u_0=u_2$, $u_0a>2$  defines the threshold for the appearance of the first excited state. As the distance $a$ increases the two eigenvalues merge [1] to one (the single eigenvalue of the independent delta potential, $E=-30.25$) in the asymmetric case $U_1=11,U_2=12$ the two levels do not merge,  instead 
they become parallel as both the levels saturate to distinct values $(-36,-30.25)$. We find that for large values of $a$ 
\begin{equation}
E_0= -m \frac{(max[U_1,U_2])^2}{2\hbar^2} \quad E_1=-m\frac{(min[U_1,U_2])^2}{2\hbar^2}.
\end{equation}
In Eq. (13), if we put $a$ very large, we get a quadratic equation for $p$ whose roots are $u_1/2$ and $u_2/2$, confirming Eq. (19). Asymmetric double well potential is often not discussed hence the question as to what the parallel  levels in Fig. 2(b) correspond to, does not arise. We would like to remark that these parallel levels  are actually the ground state eigenvalues of the independent wells of depth $U_1$ and $U_2$ (see Eq. (19)).
\begin{figure}[ht]
\centering
\includegraphics[width=5 cm,height=5 cm]{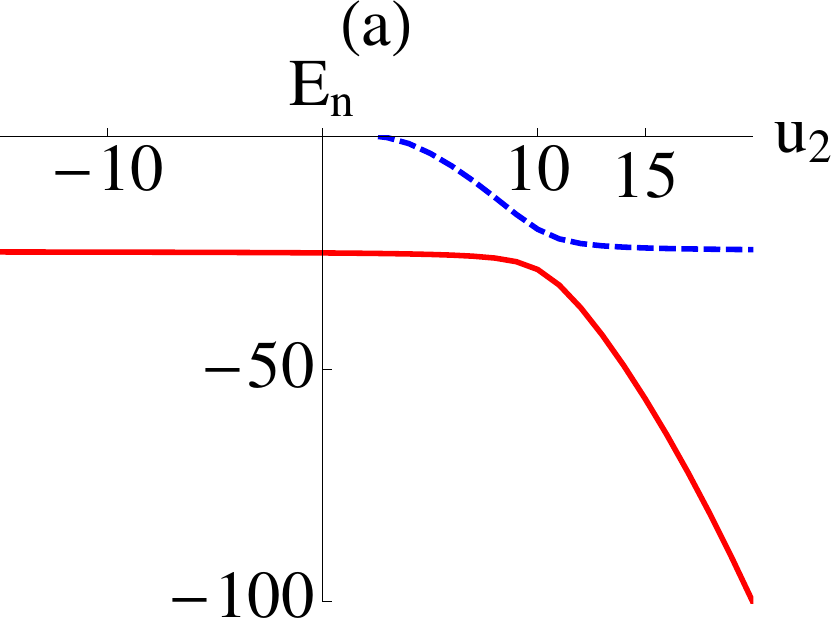}
\hskip .5 cm
\includegraphics[width=5 cm,height=5 cm]{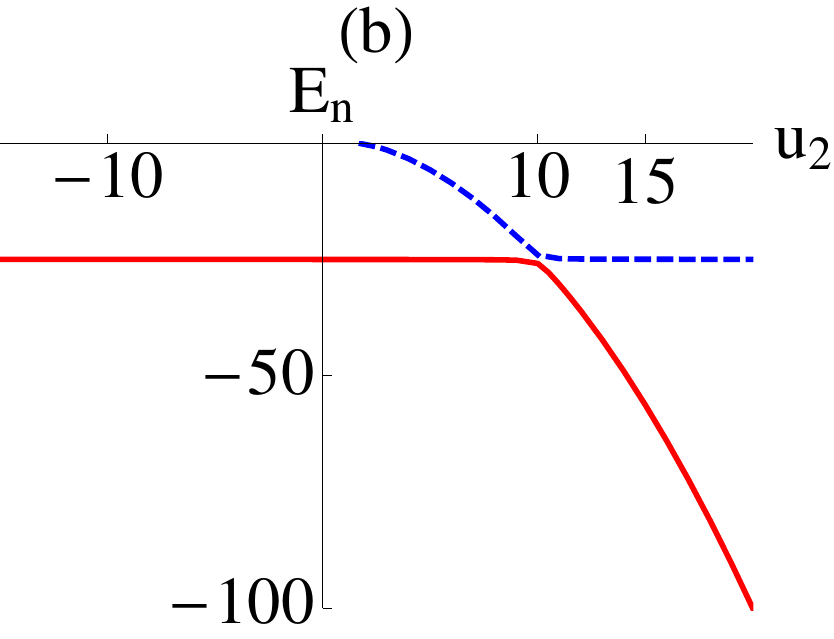}
\hskip .5 cm
\includegraphics[width=5 cm,height=5 cm]{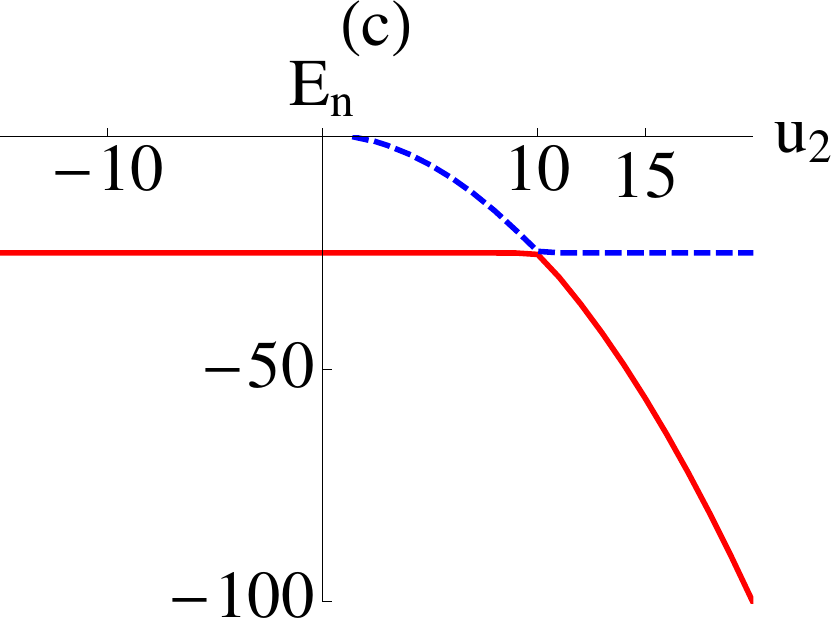}
\caption{Demonstration of avoided crossings of two levels when $u_1=10$ and $u_2$ is varied. When (a)  $a=0.5$, (b) $a=0.9$, (c) $a=1$. The special (threshold) values of $u_2$ for which $E_1$ starts appearing are 5/2, 5/4, and 10/9 for (a, b, c), respectively.}
\end{figure}

\begin{figure}
\centering
\includegraphics[width=5 cm,height=5 cm]{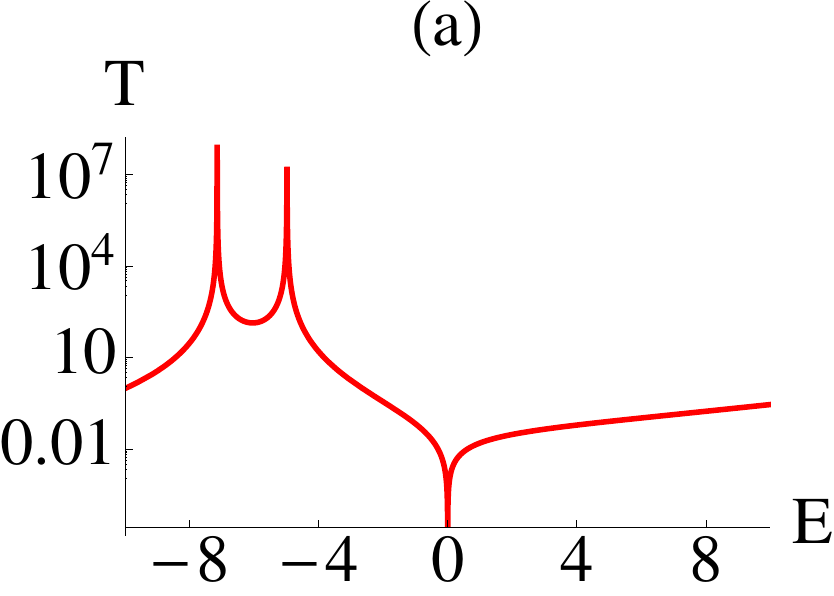}
\hskip .5 cm
\includegraphics[width=5 cm,height=5 cm]{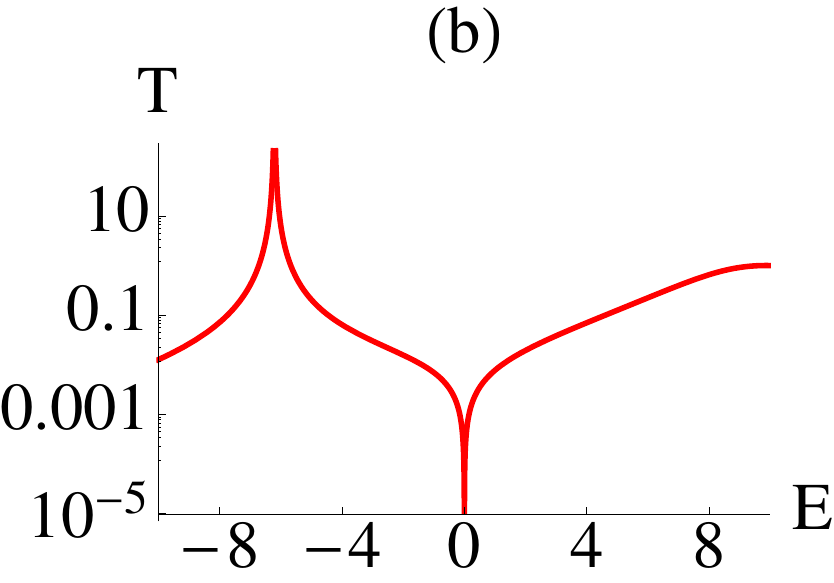}
\hskip .5 cm
\includegraphics[width=5 cm,height=5 cm]{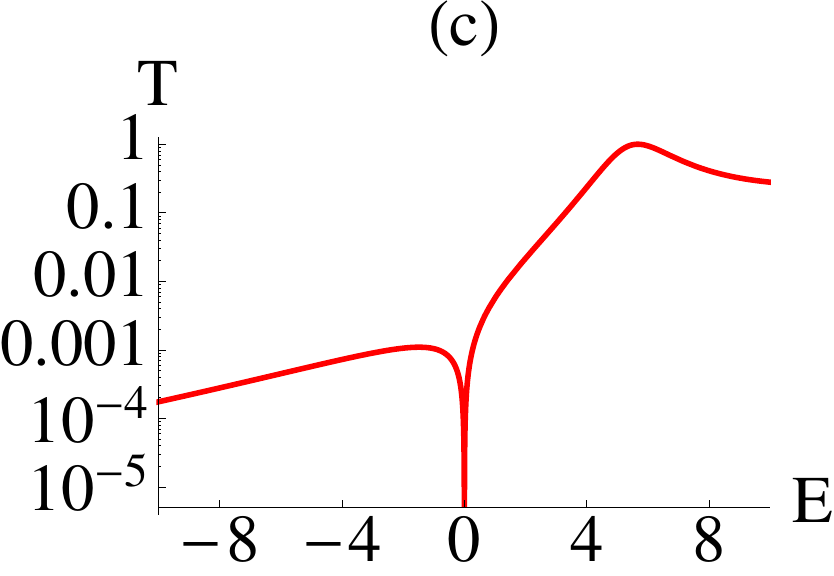}
\caption{Plots of $T(E)$ to show that negative energy poles indicate
bound states of the potential (a): $v_1=-5, v_2=-5~ (E_1=-4.98, E_0=-7.14)$, (b): $v_1=-5, v_2=5~ (E_0=-6.20)$
(c): $v_1=5, v_2=5$ (no bound states).}
\end{figure}
Further, we take $u_1=10$ and vary $u_2$ for three cases $a=0.5, 0.9 , 1 $
we see  a gradual avoided crossing of two levels solid (dashed) curves denote $E_0$  ($E_1$) (see Fig. 3). In the part (c), it is as though $E_0$ and $E_1$ have crossed however the dashed and solid nature of these curves belies the crossing. In one dimension the avoided crossing of two level though allowed is not observed usually. Rare instances have been discussed in Ref. [12], we claim that the double well potential (1) (Fig. 1(c)) presents the simplest model of avoided crossing in one dimension.

As discussed above in section 2, the common negative energy poles of $T(E)$ and $R(E)$ yield  bound states of the potential. See figure 4 that depicts two, one and no pole in $T(E)$ when the DDDP (1) possesses 2,1, and 0 bound states.

\subsection{Discrete Perfect Transmission (zero reflection) energies:}
\subsubsection{\underline {At zero energy for the attractive DDDP:}}
The reflection amplitude $\rho$ for the attractive DDDP  Fig. 1(c) deserves special attention, by changing $v_1,v_2 \rightarrow -u_1,-u_2$ in (11), we write
\begin{equation}
\rho=\frac{B}{A}=\frac{-2ik(u_1e^{-ika}+u_2e^{ika})+2iu_1u_2 \sin ka}{(2ik+u_1)(2ik+u_2)e^{-ika}-u_1 u_2 e^{ika}}.
\end{equation}
At $E=0$, $\rho$ is indeterminate ($0/0$). However, by using L'Hospital's rule, we get
\begin{eqnarray}
\rho(0)&=&0, \quad \mbox {when}~ u_1=u_0=u_2 ~\mbox {and}~u_0 a=2.\\
\rho(0)&=&\frac{u_2a(u_2a-2)}{u_2^2a^2-2u_2a+2}<1, \quad  \mbox{if}~\frac{1}{u_1}+\frac{1}{u_2}= a\quad ( u_1~ \mbox {is fixed})\\
\rho(0)&=&\frac{au_1 u_2-u_1-u_2}{u_1+ u_2-a u_1 u_2}=-1, \quad \mbox{if}  \quad \frac{1}{u_1}+\frac{1}{u_2}\ne a.
\end{eqnarray}
It may be mentioned that such cases as in Eqs. (21,22) cannot arise for $\rho$ of DDDP barrier. These Eqs. (21,22)  are new and they lead to a surprising and non-intuitive undulatory (wavelike) result that $0 \le R(0) <1$, whereas (23) is usual and most common. The result that $ 0 \le R(0) <1$ has been observed earlier and it has been called threshold anomaly [4]. In the light of the results (21,22) derived here
we conclude that this  is a critical phenomenon and  in order to bring out this critical nature of $R(0)$ graphically, in Fig. 5 we show  $R(E)$ for three cases when $u_0a=1.99,2,2.01$. Notice the dramatic result $R(0)=0$ in Fig. 5(b). In fig. 6(a,b), we show that one can arrange to have $0< R(0) <1$. Zero or small reflection  at zero energy implies that a wave packet with zero average kinetic energy, localized to one side of the potential, will spread in both directions. When the low energy components scatter against the potential, they may be transmitted but this would appear simply as wave packet spreading. 
\subsubsection{\underline {At non-zero energies in DDDP:}}
The zeros of $\rho$  in  Eq. (11) are to be obtained as 
\begin{equation}
2ik(v_1e^{-ika}+ v_2 e^{ika})+2iv_1v_2 \sin ka =0.
\end{equation}
The perfect transmission energies of a square well/barrier are known [9,13,14] to  be the eigenvalues of the corresponding  of hard box potentials. So for the square potential of width $a$ and height/depth $V_0$, the perfect transmission occurs at energies $\epsilon_n=\pm V_0 + \frac{n^2 \pi^2 \hbar^2}{2ma^2}$ which are the eigenvalues of the hard box potential of width $a$. We may  see that the aforementioned discrete energies  are also the eigenvalues of even parity states of the hard box potential of width $2a$.

We find that perfect transmission for double delta potential occurs only when it is symmetric or anti-symmetric;  further four interesting interesting cases arise here \\
{\bf Case (i):} when $v_1=-v_2=v_0$ (in fig.1(b)), we get $[4 k v_0 - 2i v_0^2] \sin ka =0$ implying $ka=n\pi$ giving
\begin{equation}
\epsilon_n=\frac{n^2 \pi^2 \hbar^2}{2ma^2},\quad n=1,2,3...
\end{equation}
the well known eigenvalues of infinitely deep well (hard-box) of width $a$.\\
{\bf Case (ii):} when $v_1=v_2= v_0$ (in Fig. 1(a))  from Eq. (24), we get
\begin{equation}
\tan k_na = - \frac{2k_n}{v_0}, \quad \epsilon_n=\frac{\hbar^2k_n^2}{2m},
\end{equation}
the roots of this equation are well known [10] as the eigenvalues of even parity states when the Dirac Delta barrier  is placed symmetrically between two rigid walls at $x=-a$ and $x=a$ (see Fig. 1(e) ).\\
{\bf Case (iii):} when $v_1=v_2=-u_0$, in this case from (24) one gets
\begin{equation}
\tan ka =\frac{2k}{u_0}, \quad \epsilon_n=\frac{\hbar^2 k_n^2}{2m},
\end{equation}
the eigenvalues [16] of the even parity states when the Dirac delta well is placed symmetrically between two rigid walls at $x=-a$ and $x=a$ (Fig.1(f)). This hard-box  potential becomes dramatically special when $u_0a=2$ (21), it is then that $E=0$ becomes the ground state eigenvalue only when the zero energy and zero-curvature solution [15,16] of the Schr{\"o}dinger equation is sought as $\psi(x)=Ax+B$. Here, we point out that this novel possibility of $E=0$ as an eigenvalue of the hard-box potential (Fig. 1(f)) forces the surprising result that  $R(0)=0$, when critically $u_0a$ becomes 2. This completes the connection of perfect transmission energies of symmetric and antisymmetric DDDP (1) with the  hard-box potentials (Fig. 1(e,f)). Next, see Figs. (7,8) displaying the phenomenon of perfect transmission when the DDDP (1) is symmetric or antisymmetric. It may be remarked that in a previous study [4] of the perfect transmission of in DDDP (1), the role of the definite parity of the potential (1) has been not been brought out. In Fig. 7, we have Dirac delta strengths as small $(\pm 5)$ and see energy oscillations in $T(E)$ whereas in Fig. 8 for higher values of the strengths ($\pm 30$), we have deep oscillations in $T(E)$. These  maxima in $T(E)$ are often misunderstood as resonances. Like the cases of square well/barriers [9,13,14], for the double Dirac delta potential, we again find perfect transmission energies $\epsilon_n$ where $T(\epsilon_n)=1$ are different from resonant energies ${\cal E}_n$ (see below). \\
{\bf Case (iv):} Other cases which are essentially non-symmetric or asymmetric, 
the roots of (24) are complex to be denoted as $E=\epsilon_n-i\gamma_n/2, \epsilon_n, \gamma_n >0$. Remarkably in these cases $T(\epsilon_n) \ne 1$ making the transmission as imperfect, see Table 1, for  asymmetric cases.
\begin{figure}[t]
\centering
\includegraphics[width=5 cm,height=5 cm]{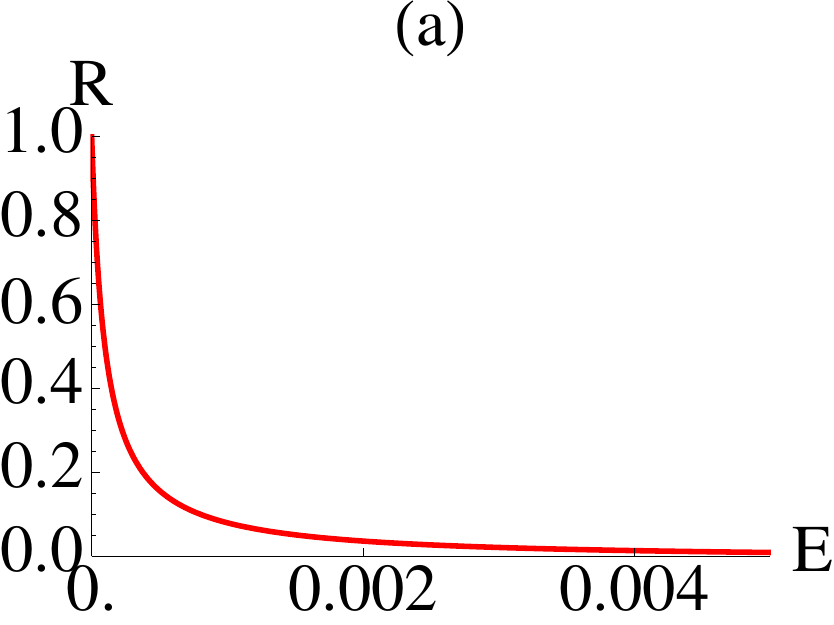}
\hskip .5 cm
\includegraphics[width=5 cm,height=5 cm]{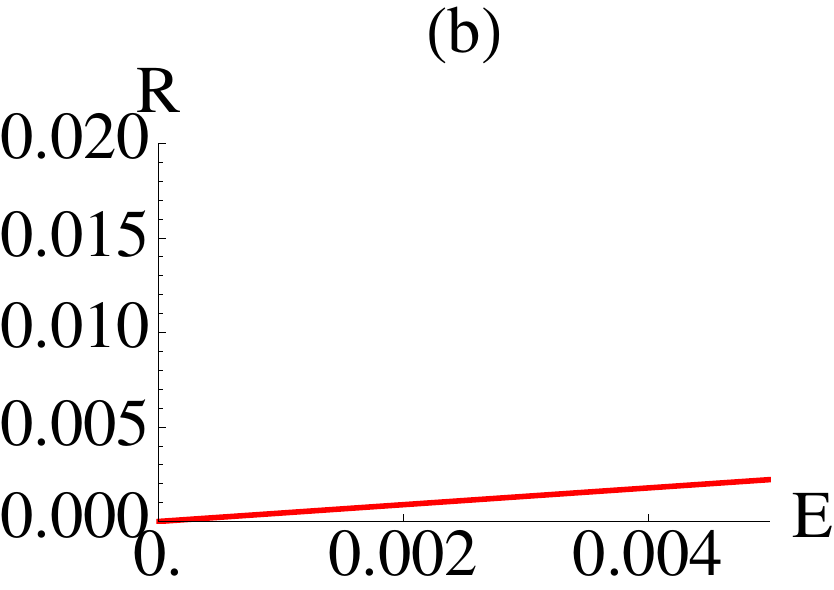}
\hskip .5 cm
\includegraphics[width=5 cm,height=5 cm]{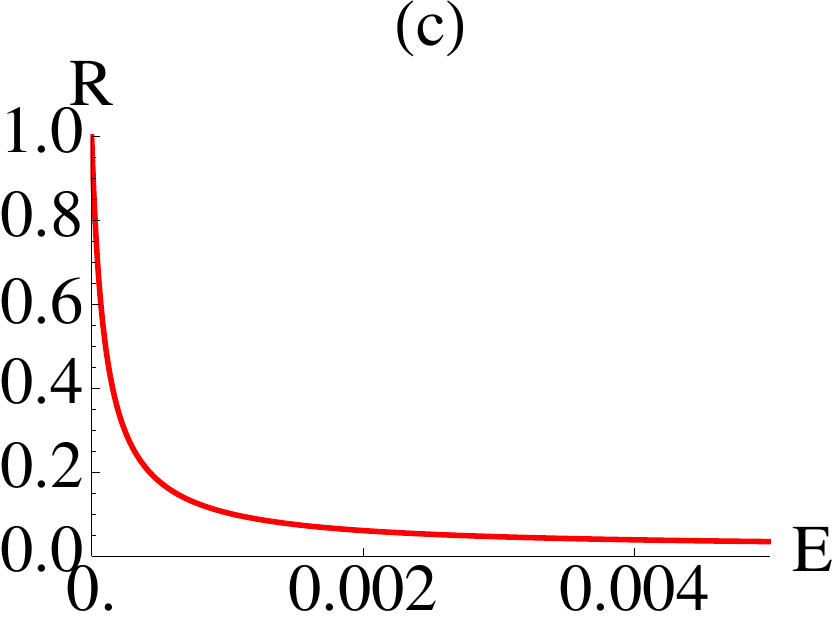}
\caption{Plot of $R(E)$, when $u_1a=u_2a=u_0a$, (a)  $u_0a=1.99$, (b) $u_0a=2$, (c) $u_0a=2.01$. Notice that $R(0)=0$ when $u_0a$ critically equals 2}
\end{figure}
\begin{figure}[t]
\centering
\includegraphics[width=5 cm,height=5 cm]{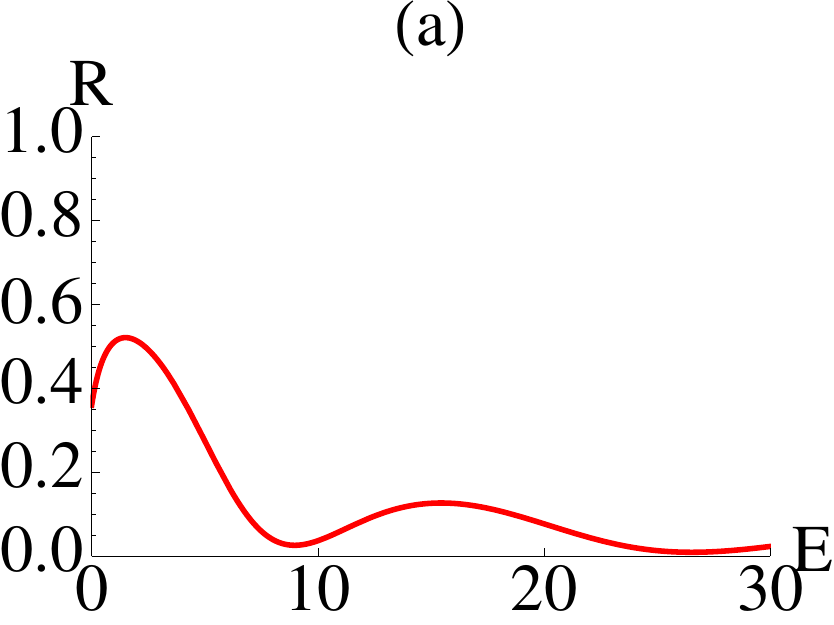}
\hskip .5 cm
\includegraphics[width=5 cm,height=5 cm]{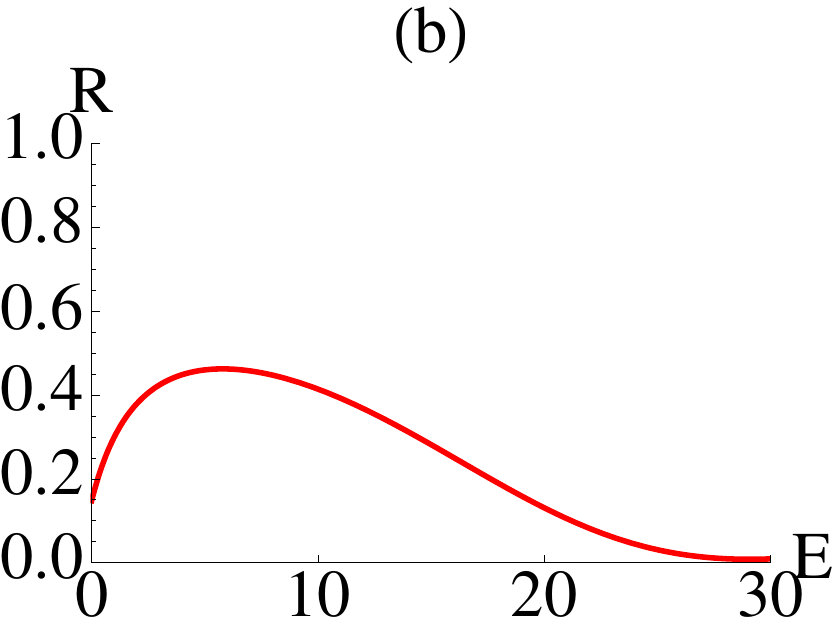}
\hskip .5 cm
\includegraphics[width=5 cm,height=5 cm]{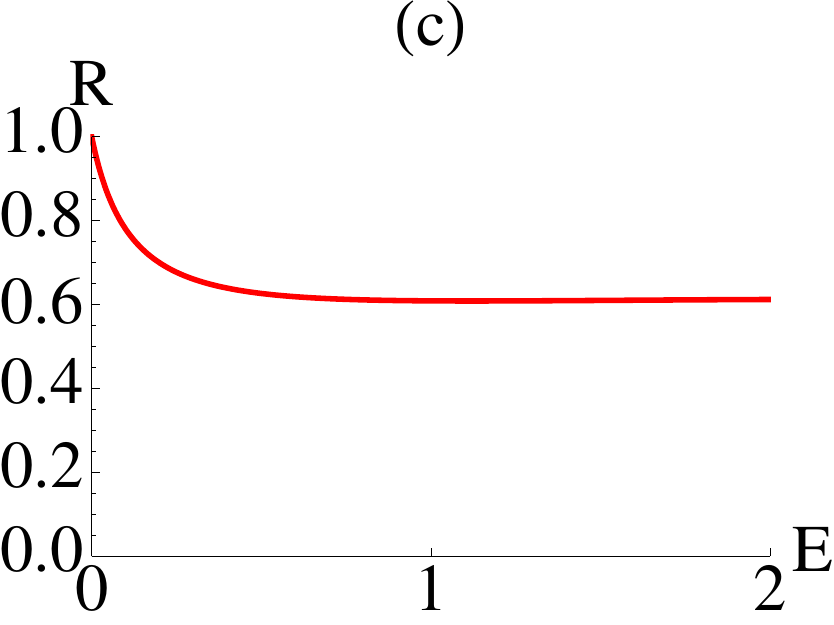}
\caption{Plot of $R(E)$, when  (a)  $u_1=2,u_2=1, a=3/2$, (b) $u_1=2, u_2=3, a=5/6$, (c) $u_1=2, u_2=3, a=1$. Notice that in (a,b) $0<R(0)<1$  when the condition $1/u_1+1/u_2=a$ is met. In (c) $R(0)=1$ is a commonly known result.}
\end{figure}

\begin{figure}[t]
\centering
\includegraphics[width=5 cm,height=5 cm]{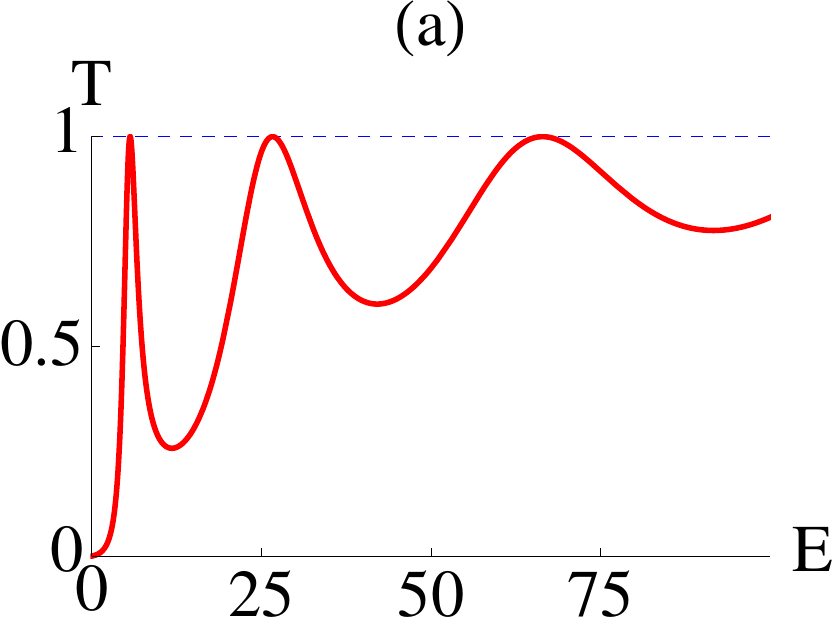}
\hskip .5 cm
\includegraphics[width=5 cm,height=5 cm]{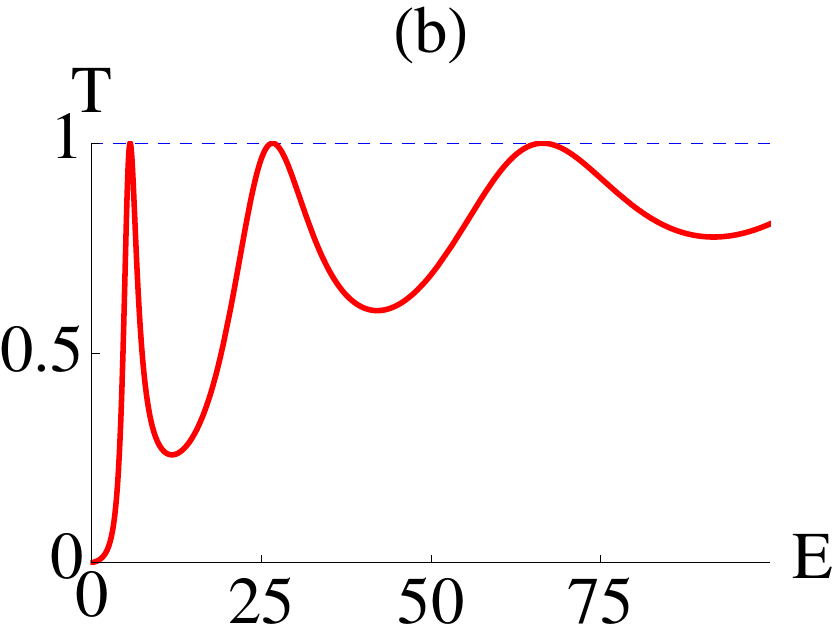}
\hskip .5 cm
\includegraphics[width=5 cm,height=5 cm]{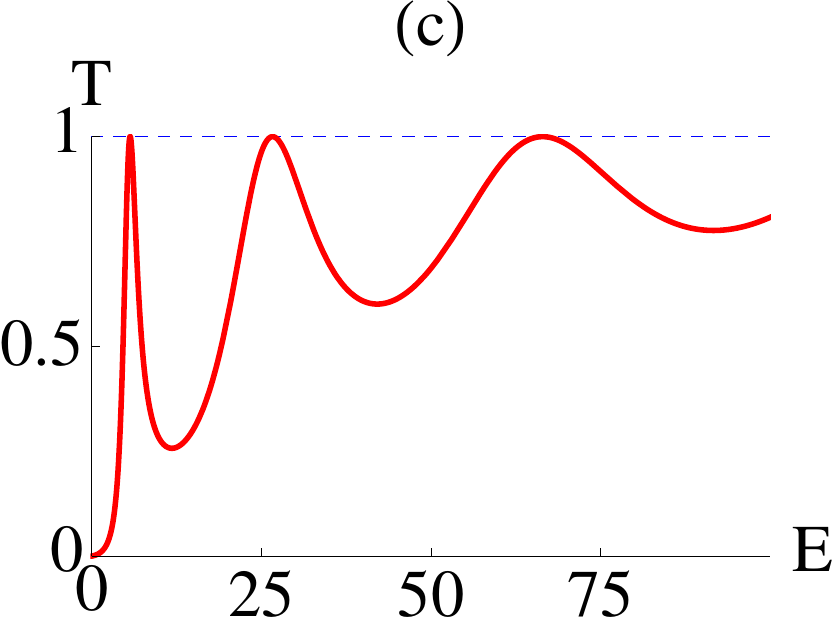}
\caption{The plot of $T(E)$ for symmetric and anti-symmetric cases when strength parameter have small values : $(a):  v_1=5, v_2 =5, (b):  v_1=5, v_2=-5, (c):  v_1=-5, v_2=-5$, $a=1$. Notice that maxima indicate perfect transmission.}
\end{figure}
\begin{figure}[ht]
\centering
\includegraphics[width=5 cm,height=5 cm]{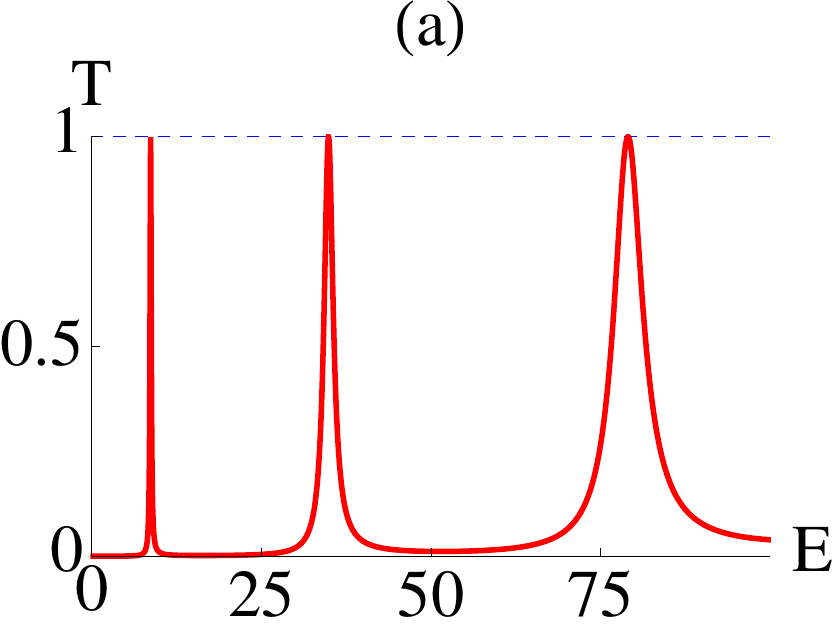}
\hskip .5 cm
\includegraphics[width=5 cm,height=5 cm]{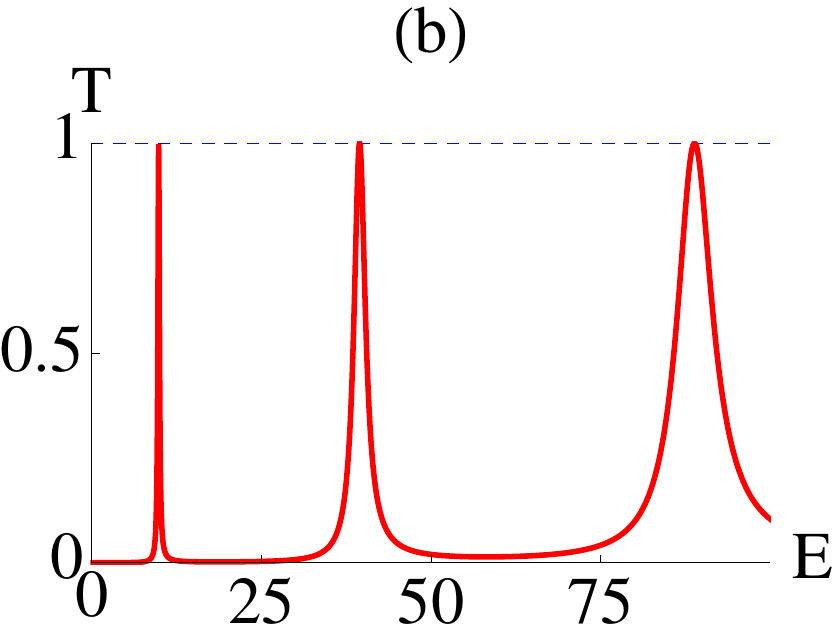}
\hskip .5 cm
\includegraphics[width=5 cm,height=5 cm]{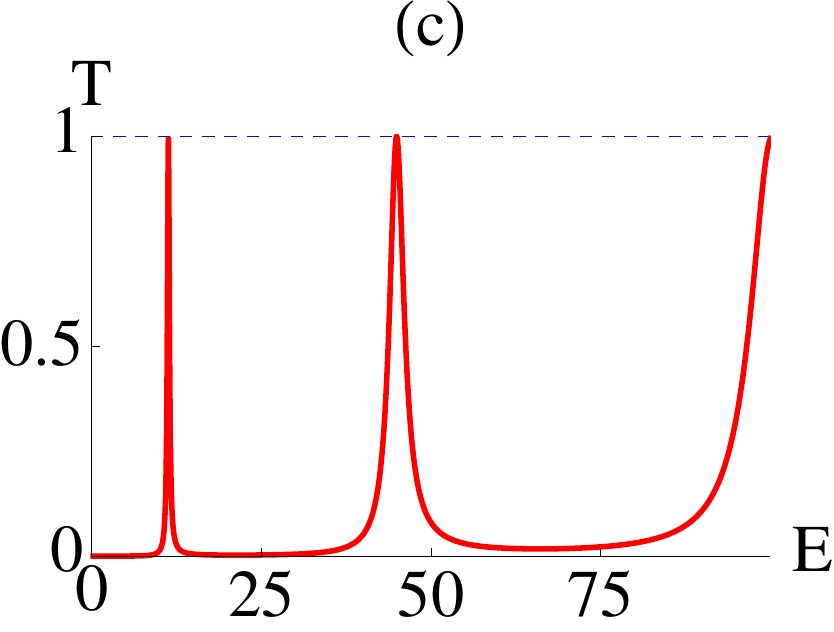}
\caption{The plot of $T(E)$ for symmetric or antisymmetric cases for large values of strength parameters : $(a): v_1=30=v_2, (b): v_1=30,  v_2=-30, (c): v_1=-30, v_2=-30$, $a=1$. Notice that maxima  indicate perfect transmission.}
\end{figure}
\subsection{Discrete complex energy resonances:}
As discussed above the complex energy resonances (Gamow's decaying states) can be obtained from the poles of $\rho$ and $\tau$ as
\begin{equation}
(2ik-v_1)(2ik-v_2)-v_1 v_2 e^{2ika}.
\end{equation}
Roots of this equation (28)  of the type $k={\cal K}_n-ik^\prime_n (E={\cal E}_n-i\Gamma_n/2), {\cal E}_n, \Gamma_n >0$ are called resonances (see Table 1) which exist in (1) whether it is a double barrier (Fig. 1(a)), a well and a barrier (Fig. 1(b)) or a double well (Fig. 1(c)) potential. The Table 1 presents first four resonances of (1) with complex discrete eigenvalues, ${\cal E}_n-i\Gamma_n/2$ for all symmetric, anti-symmetric and asymmetric cases.
Notice that resonances (unlike perfect transmission) occur whether the potential is symmetric or not and $T({\cal E}_n) \ne 1$. One may check that  ${\cal E}_n \ne \epsilon_n$.  Earlier, it has been argued [17,18] that  when $|\Gamma_n|<< {\cal E}_n$, ${\cal E}_n$ can be well approximated with $\epsilon_n$ i.e., ${\cal E}_n \approx \epsilon_n$. We would like to mention that this situation arises when the strengths of the wells or barriers are high. In this regard, the last two sections of Table 1 can be seen to support this approximation. However, in principle the Table 1 for the DDDP (1), once again [9,13,14], shows that resonance energies are different from the perfect transmission energies. 
\section{Conclusion}
We would like to conclude that our discussion of extraction of discrete eigenvalues from the scattering amplitudes/coefficients is instructive
and in that the double Dirac delta potential (DDDP) is  presented as a delightful example. It has been emphasized that discrete spectrum does not only consist of bound state eigenvalues, it also consists of resonant energies and perfect transmission energies.  
We have presented DDDP as the simplest solvable model of double wells.  The observed Avoided Crossing (AC) of two levels as the depth of one of the wells is varied slowly can be seen to be the simplest instance of this rare phenomenon of AC in one dimension. The correspondence of perfect transmission energies  with the eigenvalues of hard-box potentials (Fig.1(e,f)) is the second but more interesting instance after square well/barrier. It is now desirable to examine the generality of such a connection of perfect transmission energies of symmetric potentials  with the eigenvalues of their counterpart hard-box potentials. We also resolve that the occurrence of the surprising and non-intuitive result that $0 \le R(0) <1$ is a critical effect.  The present recourse to double Dirac delta potential provides a second example after square well/barrier to see that resonant energies and the perfect transmission energies are different and they have different origin.

\begin{table}[t]
\caption{First four resonant energies $({\cal E}_n-i\Gamma_n/2)$ and perfect transmission energies ($\epsilon_n-i\gamma_n/2$) and values of $T({\cal E}_n)$ and $T(\epsilon_n)$ for  symmetric, antisymmetric and asymmetric double Dirac delta potential (1) (Fig. 1(a,b,c)). We have taken $2m=1=\hbar^2$.}
\begin{small}
\begin{tabular}{|c||c||c||c||c|}
\hline
$~~~v_1,v_2, a$  &  I  & II & III & IV\\  
-& ${\cal E}_1-i\Gamma_1/2$, & ${\cal E}_2-i\Gamma_2/2$, &${\cal E}_3-i\Gamma_3/2$& ${\cal E}_4-i\Gamma_4/2$\\
-& $T({\cal E}_1)$ &$T({\cal E}_2)$  & $T({\cal E}_3)$&$T({\cal E}_4)$ \\
-& $\epsilon_1-i\gamma_1/2$, & $\epsilon_2-i\gamma_2/2$ &$\epsilon_3-i\gamma_3/2$& $\epsilon_4-i\gamma_4/2$\\
-& $T{(\epsilon_1)}$ &$T{(\epsilon_2)}$&$T{(\epsilon_3)}$ &$T{(\epsilon_4)}$ \\
\hline
$-3, -2.9,1$  & $15.66-9.98 i$ & $52.61-25.38 i$ & $109.90-43.37 i$ & $187.27-63.17 i$ \\ 
-&  0.9129& 0.9745& 0.9894& 0.9946\\
-& $19.25-0.14 i $& $58.73-0.25 i$&$117.95-0.36 i$ &$196.90-0.47 i$ \\
-& 0.9998 & 0.9999& 0.9999& 0.9999\\
\hline 
$-3,-3,1$ &$15.68-9.84 i$ &$52.65-25.13 i$ &$109.95-43.01 i$ &$187.33-62.70 i$\\
-&0.9134&0.9744&0.9893&0.9945\\
-&19.2074&58.6851&117.903&196.859\\
-&1&1&1&1\\
\hline
$-3,2.9,1$ &$7.82-4.74 i$ &$34.50-17.75 i$ &$81.66-34.42 i$ &$149.01-53.30 i$\\
-&0.8649&0.9599&0.9847&0.9927\\
-&$9.82-.08 i$ &$39.43-.20 i$ &$88.77-.31 i$ &$157.86-0.42 i$\\
-&0.9997&0.9999&0.9999&0.9999\\
\hline
$-3,3,1$&$7.91-4.69 i$ &$34.63-17.57 i$ &$81.81-34.12 i$ &$149.16-52.89 i$\\
-&0.8680&0.9600&0.9847&0.9927\\
-&9.8696&39.4784&88.8264&157.91\\
-&1&1&1&1\\
\hline
$3,2.9,1$ &$3.97-1.79 i$ &$21.41-11.23 i$ &$58.50-26.14 i$ &$115.81-43.91 i$\\
-&0.8655&0.9381&0.9775&0.9900\\
-&$4.70-0.04 i$ &$24.99-0.14 i$&$64.56-0.25 i$ &$123.81-0.36 i$\\
-&0.9996& 0.9999& 0.9999& 0.9999\\
\hline
$3,3,1$ &$4.01-1.77 i$ &$21.52-11.11 i$ &$58.64-25.90 i$ &$115.96-43.56 i$\\
-&0.8696&0.9387&0.9775&0.9900\\
-&4.729&25.0365&64.6169&123.867\\
-&1&1&1&1\\
\hline
$30,30,1$ &$8.68-0.10 i$ &$34.88-0.80 i$ &$78.93-2.54 i$ &$141.28-5.56 i$\\
-&0.9997&0.9992&0.9987&0.9983\\
-&8.6880&34.9042&79.0282&141.5120\\
-&1&1&1&1\\
\hline
$30,29,1$ &$8.66-0.10 i$ & $34.81-0.82 i$ &$78.80-2.61 i$ &$141.08-5.70 i$\\
-&0.9986 & 0.9982&0.9977&0.9975\\
-&$8.67-0.003 i$ &$34.83-0.002 i$&$78.90-0.07 i$ &$141.32-0.15 i$\\
-&0.9988&0.9990&0.9991&0.9993\\
\hline
\end{tabular}
\end{small}
\end{table}

\end{document}